\documentclass[aps,prr,twocolumn,superscriptaddress,preprintnumbers,showkeys]{revtex4-2}

\usepackage{amsmath,amssymb,amsfonts}
\usepackage{bm}
\usepackage{graphicx,color}
\usepackage{verbatim}
\usepackage{float}
\usepackage{dcolumn}
\usepackage{natbib}
\usepackage{hyperref}

\newcommand{\ie}{\emph{i.e.}}

\newcommand{\ang}[1]{\left\langle #1 \right\rangle}

\begin{document}

\title{True scale-free networks hidden by finite size effects}

\author{Matteo Serafino}
\affiliation{IMT School for Advanced Studies, 55100 Lucca, Italy}
\author{Giulio Cimini}
\affiliation{Physics Department and INFN, University of Rome Tor Vergata, 00133 Rome (Italy)}
\affiliation{IMT School for Advanced Studies, 55100 Lucca, Italy}
\affiliation{Institute for Complex Systems (CNR) UoS Sapienza, 00185 Rome (Italy)}
\author{Amos Maritan}
\affiliation{Department of Physics, University of Padova, 35131 Padova, Italy}
\author{Andrea Rinaldo}
\affiliation{Laboratory of Ecohydrology, \'{E}cole Polytechnique F\'{e}d\'{e}rale de Lausanne, CH-1015 Lausanne, Switzerland}
\affiliation{Department of Civil, Constructional and Environmental Engineering, University of Padova, 35131 Padova, Italy}
\author{Samir Suweis}
\affiliation{Department of Physics, University of Padova, 35131 Padova, Italy}
\author{Jayanth R. Banavar}
\affiliation{Department of Physics and Institute for Fundamental Studies, University of Oregon, Oregon 97403, USA}
\author{Guido Caldarelli}
\affiliation{Department of Molecular Sciences and Nanosystems (DSMN), Ca' Foscari University of Venice, 30172 Venezia Mestre, Italy}
\affiliation{European Centre for Living Technology, 30124 Venice, Italy}
\affiliation{Institute for Complex Systems (CNR) UoS Sapienza, 00185 Rome (Italy)}
\affiliation{London Institute for Mathematical Sciences, W1K2XF London, United Kingdom}


\begin{abstract}
We analyze about two hundred naturally occurring networks with distinct dynamical origins to formally test whether the commonly assumed hypothesis of an underlying scale-free structure is generally viable. This has recently been questioned on the basis of statistical testing of the validity of power law distributions of network degrees by contrasting real data. Specifically, we analyze by finite-size scaling analysis the datasets of real networks to check whether purported departures from the power law behavior are due to the finiteness of the sample size. In this case, power laws would be recovered in the case of progressively larger cutoffs induced by the size of the sample. We find that a large number of the networks studied follow a finite size scaling hypothesis without any self-tuning. This is the case of biological protein interaction networks, technological computer and hyperlink networks, and informational networks in general. Marked deviations appear in other cases, especially infrastructure and transportation but also social networks. We conclude that underlying scale invariance properties of many naturally occurring networks are extant features often clouded by finite-size effects due to the nature of the sample data.
\end{abstract}
\keywords{network form and function, degree distribution, power laws, finite size scaling, statistical physics}

\maketitle

Networks play a vital role in the development of predictive models of physical, biological, and social collective phenomena \cite{barabasi2016network,caldarelli2007scale,cimini2019review}.
A quite remarkable feature of many real networks is that they are believed to be approximately scale-free: 
the fraction of nodes with $k$ incident links (the degree) follows a power law $p(k)\propto k^{-\lambda}$ for sufficiently large value of $k$ \cite{barabasi1999emergence,klarreich2018scant}. 
The value of the exponent $\lambda$ as well as deviations from power law scaling provides invaluable information on the mechanisms underlying the formation of the network such as small degree saturation, variations in the local fitness to compete for links, and high degree cut-offs owing to the finite size of the network. 
Indeed real networks are not infinitely large and the largest degree of any network cannot be larger than the number of nodes. 
Finite size scaling \cite{fisher1967theory,stanley1971introduction,binder1992montecarlo,kim1996numerical,stanley1999scaling,redner2001guide,corral2016exact}, firstly developed in the field of critical phenomena and renormalization group, is a useful tool for analyzing deviations from pure power law behavior as due to finite size effects. 
Here we show that despite the essential differences between networks and critical phenomena, finite size scaling provides a powerful framework for analyzing the scale-free nature of empirical networks. 

The search of ubiquitous emergent properties occurring in several different systems and transcending the specific system details is a recurrent theme in statistical physics and complexity science \cite{goldenfeld1999simple}. Indeed the presence and the type of such ``universal” law gives insights on the driving processes or on the characteristic properties of the observed system. Notably, complex systems have the propensity to display ``power law” like relationship in many diverse observables (such as event sizes and centrality distribution, to name a few). In particular the power law shape of the degree distribution, which is the hallmark of {\em scale-free} networks, leads to important emergent attributes such as self-similarity in the network topology, robustness to random failures and fragility to targeted attacks. Notably scale invariance extends far beyond the degree distribution, affecting many other quantities as  weighted degree, betweenness \cite{barrat2004architecture} and degree-degree distance \cite{zhou2019powerlaw}.

In the last decade the existence of such power laws in complex networks (but also in other areas \cite{Stumpf665}, e.g., law in language \cite{moreno2016large}) has been questioned \cite{clauset2009power}. A reason of the shift in such conclusion is in the availability of larger (and new) datasets, and especially in improved statistical methods. Recently, Broido and Clauset\cite{broido2018scale} fitted a power law model to the degree distribution of a variety of empirical networks and suggested that scale-free networks are rare. Voitalov{\em et al.}\cite{voitalov2018sfnwelldone} rebutted that scale-free networks are not as rare if deviations from pure power law behavior are permitted in the small degree regime. The different conclusions may depend on very fine but critical assumptions at the basis of the statistical test for the power law hypothesis. Moreover, a crucial point that is typically ignored but represents the condition for the proper use of maximum likelihood methods is the independence of the empirical observations \cite{gerlach2019testing}. In this work we tackle the problem of detecting power laws in networks from a different perspective, based on the the machinery of finite size scaling.

\begin{figure*}[ht!]
\centering
\includegraphics[width=0.75\textwidth]{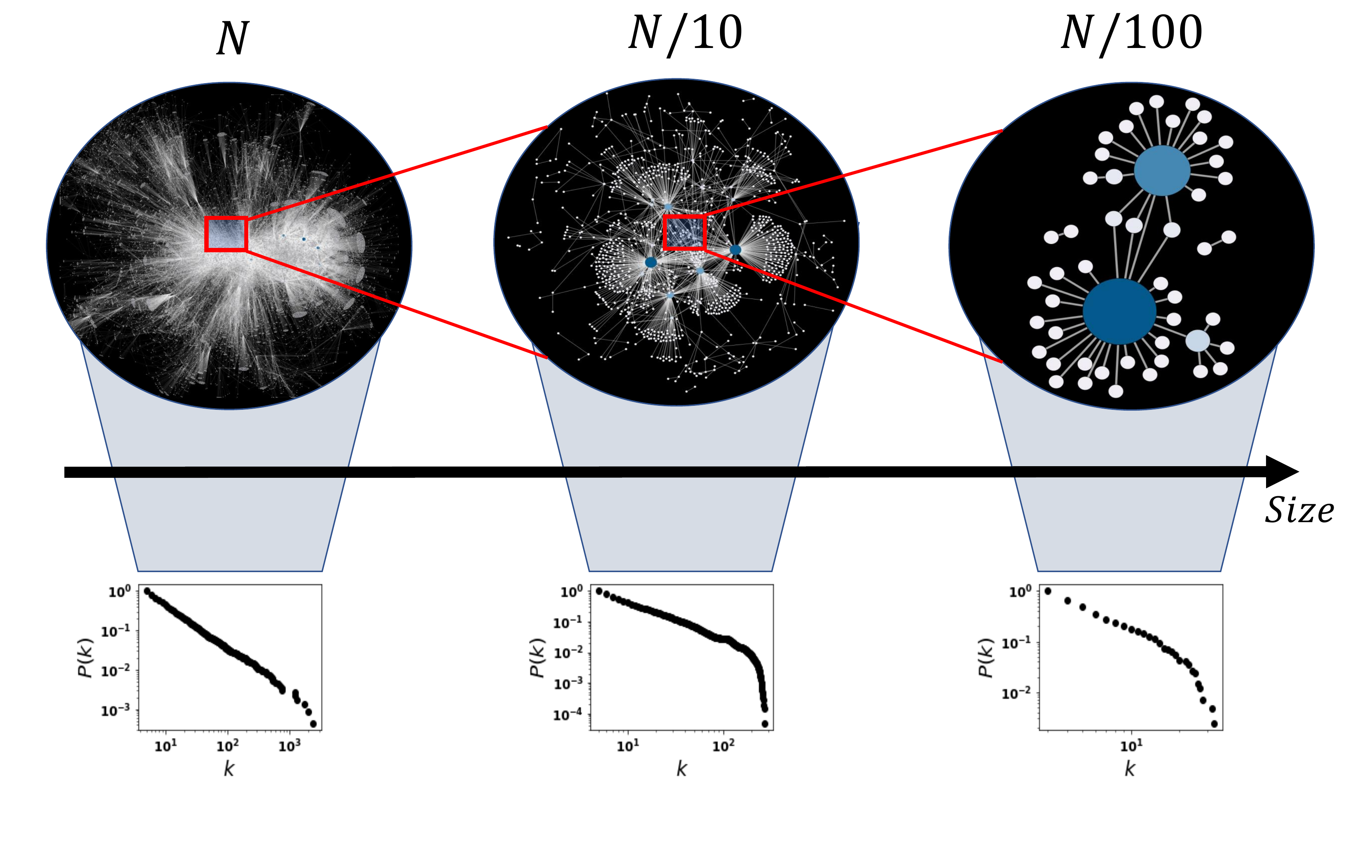}
\caption{An illustrative example of the concept of scale invariance in a network. If the degree distribution of the network is scale-free, then small sub-samples of the network will have the same distribution -- i.e., the degree structure of the network will not be altered (apart from small deviations at low $k$). The network in this example is a snapshot of the structure of the Internet at the level of autonomous systems \cite{newman_2006}}
\label{fig:0}
\end{figure*}

Statistical physics of critical phenomena teaches us that a system at criticality exhibits power law singularities of physical quantities such as, for example, the compressibility, the specific heat, the density difference between the liquid and vapor, as well as the latent heat. Water at its critical point exhibits fluctuations at all scales between the molecular length scale and the size of the container, which could be macroscopically large. Moreover, one finds thoroughly mixed droplets of water and bubbles of gas. Indeed, any large part of the system looks like the whole – the system is self-similar. The length scale of these droplets and bubbles extends from the molecular scale up to the correlation length, which is a measure of the size of the largest droplet or bubble. The divergence of the correlation length in the vicinity of a phase transition at the thermodynamic limit thus suggests that properties near the critical point can be accurately described within an effective theory involving only long-range collective fluctuations of the system. However, both in experiments and in numerical simulations, the infinite size limit cannot be reached and thus one observe deviations from the predicted thermodynamics limit behavior. The finite size scaling (FSS) ansatz has been developed precisely to infer the singular behavior (i.e., the exponents determining the universality classes) of the physical properties of a system in the thermodynamic limit, having only information on the system properties at finite sizes.

FSS has yet a more general validity and does not require the existence of a phase transition or an evolution process. Indeed, even though it was initially used to study finite systems near the critical point of the corresponding infinite system, FSS can be actually applied to describe structures that are self-similar when observed in a certain range of scales. As an example, we consider a Cantor set where we stop the procedure to divide intervals in three parts and removing the middle one at a scale $s_0=3^{-m}$. This corresponds to a fractal structure on scales between $s_0$ and 1, and to a non-fractal structure on scale smaller than $s_0$. If we measure the total length, $L(s)$, of the set with a stick of length $s=3^{-n}$ we find $L(s)=s^{1-D} F(s/s_0)$ where $F(x)=1$ when $x>1$ whereas $F(x)=x^{1-D}$ when $x<1$ and $D=\log_3 2$ is the Hausdorff–Besicovitch (or fractal) dimension of the Cantor set. Another illustration of FSS analysis is given by the truncated geometrical series $S(x,N)=\sum_{0}^{N-1}x^n$. When $x$ is close to 1 it is easy to see that $S(x,N)=t^{-1}F(tN)$, where $t=1-x$ and $F(z)=1-e^{-z}$. As a matter of fact, the FSS approach has been used to test scale invariance (and self similarity) also for non-critical systems such as (just to mention some very famous examples) polymers in confined geometries \cite{de1979scaling} and interfaces \cite{safran2018statistical,nelson2004statistical}. 
In view of the above, FSS can also implemented on the well-established models of scale-free networks-like the Barab\'asi-Albert model where the scale free behavior is not an emergent property at a critical point. Whether or not the same hypotheses holds for real world network does not undermine the possibility of applying FSS to them.

Employing the FSS machinery to test whether empirical networks display scale-free behavior in their degree distribution is not straightforward though. Unlike for physical systems, representations of a network at different scales are typically not available. Thus in order to test whether a network shows a power law distribution of the degree, we construct by hand smaller sized representations of it in an unbiased manner. We then use the characteristics of the large original network as well as the derived sub-networks to test the scale-free hypothesis. Figure \ref{fig:0} shows an illustration of this procedure for a snapshot of the structure of the Internet at the level of autonomous systems \cite{newman_2006}. Subsection A provides a brief summary of finite size scaling applied to network topology. 
Subsection B presents an independent method of determining whether networks are scale-free based on analyses of the size dependence of the ratio of moments of the degree distributions. 
Subsection C provides information on the sampling scheme used to build sub-networks and on the region selected for the scaling analysis.

In the Results section we test the scale-free hypothesis (intended as the power law behavior in the degree distributions) on around two hundred large empirical networks (those considered in \cite{broido2018scale} and \cite{voitalov2018sfnwelldone}). Remarkably, we find that such a venerable hypothesis cannot be rejected for many (but not all) networks. Moreover the two scaling exponents for such networks satisfy an additional scaling relationship, which derives from the shape of the degree cross-over in scale-free networks. We benchmark our results against the quality measure of the well-known scale-free graph introduced by Barab\'asi and Albert \cite{barabasi1999emergence}. Further we show that finite size scaling allows discerning pure power laws from log-normal and Weibull distributions. In conclusion, our results support the claim that scale invariance is indeed a feature of many real networks, with finite size effects accounting for quantifiable deviations. 

\subsection*{A. Finite Size Scaling of networks}
\label{section 1}

A scale-free network is postulated to have a degree distribution $p(k)\propto k^{-\lambda}$ beyond some lower degree cut-off $k_{min}$. 
For an infinitely sized network, since $k_{min} \geq 1 $, the exponent $\lambda > 1$ in order for $p(k)$  to be normalizable. 
In what follows, we will consider the cumulative distribution $P(k)=\int_{k}^{\infty}p(q)dq\propto k^{-\gamma}$ where $\gamma =\lambda-1>0$.

Networks are of course not infinitely large. In a network comprising $N$ nodes, $k$ can be at most equal to $N-1$. This is the intrinsic limit on $k$ given by the network size. 
Thus it is plausible that, below some $k_{c}$ (cross-over value), the degree distribution follows a power law behavior as would be expected 
for an infinite network but falls more rapidly beyond $k_{c}$. The finite size scaling hypothesis states that
\begin{equation}
P(k,N) = k^{-\gamma}f(kN^{d})
\label{scaling}
\end{equation}
where $d<0$. The remarkable simplifying feature of the scaling hypothesis is that $P$ is not an arbitrary function of the two variables $k$ and $N$ but rather $k$ and $N$ combine in a non-trivial manner 
to create a composite variable. The behavior of the system is fully defined by the two exponents, $\gamma$ and $d$, and the scaling function $f$. 
The exponent $d<0$ so that, for an infinite size network ($N \to \infty$), the argument of $f$ approaches zero. 
A pure power law decay of $P(k,N)$ with $k$ for very large $N$ requires that $f(x) \to$ constant as $x \to 0$. The additional normalization condition is $f(x)\to 0$ sufficiently fast when $x\to 1$. The finite size effects are quantified by the behavior of the function $f$ as its argument increases, e.g., when $k \gtrsim k_c$. For a network with a finite number of nodes, the degree distribution does not follow a pure power law but is modified by the function $f$ (see also \cite{Krapivsky_2002} for a discussion of finiteness in the context of growing network models).

A powerful way of assessing whether a network is scale invariant is to confirm the validity of the scaling hypothesis and determine the two exponents and the scaling function $f$ by using the collapse plot technique. 
One may recast Eq. (\ref{scaling}) as
\begin{equation}
P(k,N) k^{\gamma}=f(kN^{d}).
\label{scaling2}
\end{equation}
Then the path forward is simple. For networks belonging to the same class but with different $N$, one optimally selects two fitting parameters $\gamma$ and $d$ by seeking to collapse plots of $P(k,N)k^{\gamma}$ versus $kN^d$ for different $N$ on top of each other \cite{bhattacharjee2001measure}. The fidelity of the collapse plot provides a measure of self-similarity and scale-free behavior, the optimal parameters are the desired exponents, and the collapsed curve is a plot of the scaling function.

We start out with a single representation of an empirical network with $N$ nodes. For purposes of the scaling collapse plot, we seek additional representative networks of smaller sizes. In order to accomplish this, we obtained the mean degree distributions of multiple sub-networks of sizes $\frac{N}{4}$, $\frac{N}{2}$ and $\frac{3N}{4}$, which were then collapsed on to each other and the original network to create a master curve.
The quality $S$ of the collapse plot is then measured as the mean square distance of the data from the master curve in units of standard errors. $S$ is thus like a reduced $\chi^2$ test, and should be around one if the data really collapse to a single curve and much larger otherwise 
\cite{houdayer2004low}.

Note that as a measure of the size of a network (or sub-network), one may use the number of nodes $N$ or alternatively the number of links $E$. The scaling function in this case reads as follows:
\begin{equation}
P(k,E)k^{\gamma}=f_{E}(kE^{d_{E}}), 
\label{scalingedge}
\end{equation}
where the exponent $\gamma$ is the same as before and the exponent $d_{E}<0$ ought to be equal to the previously introduced exponent $d$ for networks satisfying the finite size scaling hypothesis (see next section). 

\subsection*{B. Ratio of moments test}
\label{section 2}
A simple alternative and independent test of the scale-free hypothesis is to study the size dependence of the ratio between the $i$-th and the $(i-1)$-th moments of $k$, for various $i$. 
The $i$-th moment $\ang{k^i}$ is defined to be
\begin{equation}
\ang{k^i}=\int_{k_{min}}^{\infty} dk\,k^{i-1} k^{-\gamma} f(kN^{d})\propto N^{-d(i-\gamma)}
\label{moments}
\end{equation}
provided $i>\gamma$. Instead if $i \leq \gamma$, $\ang{k^i}$ converges to a constant value for $N \to \infty$. Therefore when $i-1 > \gamma$,
\begin{equation}
\ang{k^i} \Big/ \ang{k^{i-1}}\propto N^{-d},
\label{constant}
\end{equation}
independently of $i$. Thus, for a scale-free network, a log-log plot of the ratio of consecutive moments versus $N$ is a straight line with slope $-d$. Likewise
\begin{equation}
\ang{k^i}=\int_{k_{min}}^{\infty}  dk\,k^{i-1} k^{-\gamma} f_E(kE^{d_E})\propto E^{-d_E(i-\gamma)}
\label{momentsE}
\end{equation}
when $i>\gamma$, otherwise $\ang{k^i}$ goes to a constant for $E \to \infty$. Therefore when $i-1 > \gamma$,
\begin{equation}
\ang{k^i} \Big/ \ang{k^{i-1}}\propto E^{-d_E}.
\label{constant2}
\end{equation}
The exponents $d$ and $d_E$ are not independent for scale-free networks. On the one hand, equations (\ref{moments}) and (\ref{momentsE}) imply $E\propto N^{d/d_E}$. On the other, in general $\ang{k}\propto E/N\propto N^{d/d_E-1}$. Due to the above equations $\ang{k}$ is constant for scale-free networks with $\gamma>1$, implying that $d=d_E$. 
Thus the difference between $d$ and $d_{E}$ values (that we statistically assess through their $Z$-score) provides an independent quality measure of the scale-free attributes of a network. 

\subsection*{C. Sub-sampling and scaling region}
\label{section 3}
In order to generate a sub-network of a given size $n<N$, we pick $n$ nodes at random among the $N$ nodes of the original network, removing all the other nodes and the links originating from them. It is well known that the sub-sampling procedure modifies the shape of the degree distribution of the network. In particular, sub-networks of scale-free networks are not scale-free because of deviations at low $k$ values \cite{stumpf2005subsets} (this happens independently of the sampling scheme adopted \cite{lee2006statistical}). The problem of the left tail of the distribution however applies more generally, because deviations from the scale-free behavior at low degrees are rather common in empirical and network models. 
Therefore we perform the scaling analysis described in subsections A and B only for $k\ge k_{min}$, where the lower bound of the scaling region $k_{min}$ is chosen such that the empirical distribution of the original network and its best power law fit (with exponent $\Gamma$, computed with the maximum-likelihood method of Clauset, Shalizi and Newman \cite{clauset2009power}, see Methods) are as similar as possible above $k_{min}$ 
\cite{alstott2014powerlaw}. 
In the Supplementary Information we show that this allows us to get rid of any deviations induced by the sub-sampling scheme. However, when the empirical distribution of the network deviates substantially from a power law over its entire domain, then the estimated $k_{min}$ can become very large and may even diverge. In these cases the number of nodes $n^*$ of the (sub-)network with $k\ge k_{min}$ becomes very small or vanishing, yielding an unstable or undefined collapse. 
We thus use $n^*\ge\ln N$ as a condition on as the minimum number of nodes in each (sub-)network for the feasibility of the scaling analysis. 

\section*{Results}
To sum up, two independent statistical tests of the scale-free attributes of a network explained in subsections A and B are the quality of the collapse $S$ (\ie, the reduced $\chi^2$ between data and master curve) and the compatibility of $d$ and $d_E$ (measured through their $Z$-score). Figure S1 in the Supplementary Information outlines the flow of the analysis.
In line with Broido \& Clauset \cite{broido2018scale} and Voitalov {\em et al.} \cite{voitalov2018sfnwelldone}, we use these tests to define a classification for the degree distribution of empirical networks:
\begin{itemize}
\item\textbf{SSF} (strong scale-free) if $S\leq 1$ and $Z_{dd_{E}}\leq1$,
\item\textbf{WSF} (weak scale-free) if $S\leq 3$ and $Z_{dd_{E}}\leq3$,
\item\textbf{NSF} (non scale-free) otherwise or when $n^*<\ln N$ for the original network or any of its sub-networks.
\end{itemize}
Note the nestedness of the classification, for which a SSF network is also WSF.

\begin{figure*}[ht!]
\begin{center}
\includegraphics[width=0.8\linewidth]{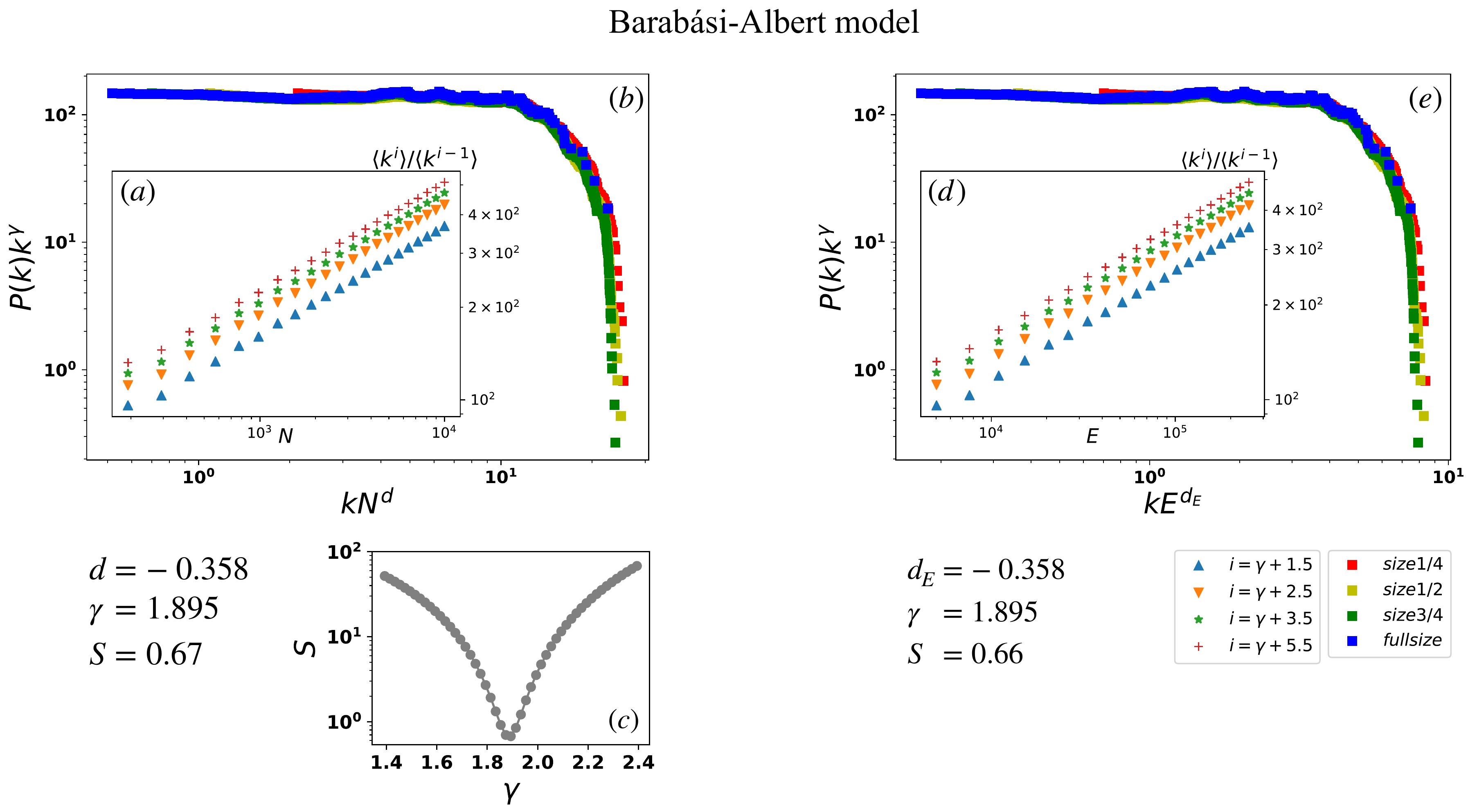}
\caption{Scaling analysis on a numerical realization of the Barab\'asi-Albert model. 
The network has $N=10^4$ nodes and the minimum node degree is $k_{min}=14$. The best power law fit on this network yields $\Gamma=1.89\pm0.02$. Note this value is smaller than $\Gamma=2$ because of deviations from the pure power law at small $k$s: indeed, the theoretical $P(k)$ in the Barab\'asi-Albert model goes as $[k(k + 1)(k + 2)]^{-1}$  \cite{dorogovtsev20004structure}). Panels (a), (b), (c) show results of the scaling analysis using the number of nodes as for Eqs. (\ref{scaling2}) and (\ref{constant}). Inset (a) reports the dependence of various moment ratios on $N$; fitting these slopes yields $d=-0.358\pm0.035$. The main panel (a) shows the collapse of the cumulative degree distributions when scaled with $N$. The best collapse is obtained with $\gamma=1.89\pm0.06$ and yields $S=0.67$. Panel (c) shows how the quality of the collapse reported in (a) varies on moving away from the optimal value of $\gamma$. Panels (d), (e) further show results of the scaling analysis using the number of links as for Eqs. (\ref{scalingedge}) and (\ref{constant2}). In this case, the moment ratio test of inset (d) returns $d_E=-0.351\pm0.031$ while the best collapse of the cumulative degree distributions reported in the main panel (e)  is obtained with $\gamma=1.89\pm0.05$ and yields $S=0.66$.}
\label{Barabasimodel}
\end{center}
\end{figure*}

\begin{figure*}[ht!]
\begin{center}
\includegraphics[width=0.45\linewidth]{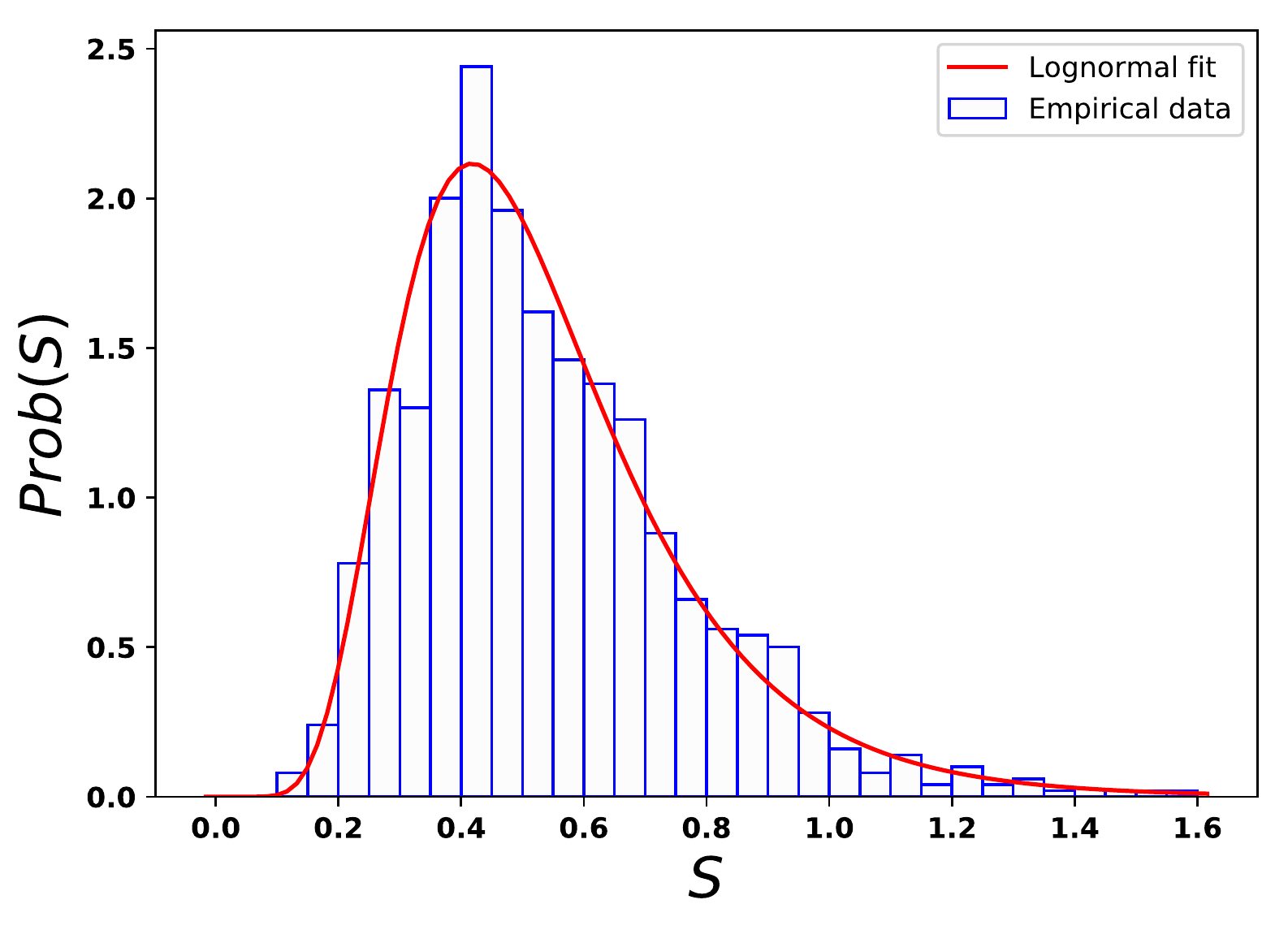}
\caption{Empirical distribution of the quality of collapse $S$ obtained from finite size scaling analysis on 1000 realizations of the Barab\'asi-Albert graph (same parameters of Figure \ref{Barabasimodel}). 
The distribution is well fitted by a log-normal with $\mu=-0.70 \pm 0.1$ and $\sigma=0.414 \pm 0.009$.}
\label{Barabasimodel_stats}
\end{center}
\end{figure*}

\subsection*{Power law and Poisson distribution}

We start analyzing the reference cases of Barab\'asi-Albert \cite{barabasi1999emergence} and Erd\H{o}s-R\'enyi \cite{erdos1959random} models whose behavior is known. In the former case $p(k)\sim k^{-3}$, whereas, in the latter case $p(k)\sim\mbox{Poisson}_{\bar{k}}(k)$.  
Figure \ref{Barabasimodel} shows that for a realization of the Barab\'asi-Albert graph the degree distributions of the (sub-)networks result in a collapse of very high quality. The power law exponent $\gamma$ yielding the best collapse is consistent with the value $\Gamma$ obtained by maximum-likelihood fitting the degree distribution of the mother network with a power law \cite{clauset2009power}. Additionally, the moments ratio are indeed parallel lines, with compatible slopes $d$ and $d_E$. 
A more robust statistics is obtained by analysing $1000$ realizations of the Barab\'asi-Albert model (Figure \ref{Barabasimodel_stats}). Within this sample, $98\%$ of the networks are classified as SSF while $2\%$ as WSF. The estimated scaling exponents are all consistent with each others among the different realizations.

For the Erd\H{o}s-R\'enyi model the estimated $k_{min}$ for the degree distribution is so large that it is not possible to have (sub-)networks with number of nodes $n^*\ge\ln N$ (in principle, for this network, the $k_{min}$ estimated from the KS test should be larger than the largest degree of the network). As such, the Erd\H{o}s-R\'enyi graph is classified as NSF. We obtained the same outcome in an ensemble of 1000 realization of this network model.

\subsection*{Alternative fat tail distributions}

While the power law is the only distribution featuring scale invariance, there are other distributions characterized by a fat right tail that can resemble a power law in finite systems. Hence determining which of these distribution better fits empirical network data is often a nontrivial task. In particular the classical approach based on $p$-values computed from a Kolmogorov-Smirnov test (see Methods) is able to rule out some competing hypothesis but not to confirm one \cite{clauset2009power}. Moreover, the hypothesis testing approach may fail when applied to regularly varying distributions \cite{voitalov2018sfnwelldone}. It is therefore meaningful to put our finite size scaling approach to the test of alternative fat tail distributions. Here we consider the representative cases of the log-normal and Weibull distributions. 
The log-normal distribution 
$p(\ln k)=\mbox{Normal}(\mu,\sigma)$ is characterized by parameters $\mu$ and $\sigma$, respectively the mean and standard deviation of the variable's natural logarithm. For large values of $\sigma$ this distribution is highly skewed and features a fat tail for large $k$ values. 
The Weibull distribution $p(k)=(h/l^h)k^{h-1} \exp\left[-(k/l))^h\right]$ is characterized by parameters $h$ (shape) and $l$ (scale). The fat tail in this case appears for $h \to 0$.
We use the Viger-Latapy algorithm \cite{viger2005efficient} to generate networks with these degree distributions.

\begin{figure*}[p!] 
\begin{center}
\includegraphics[width=0.8\linewidth]{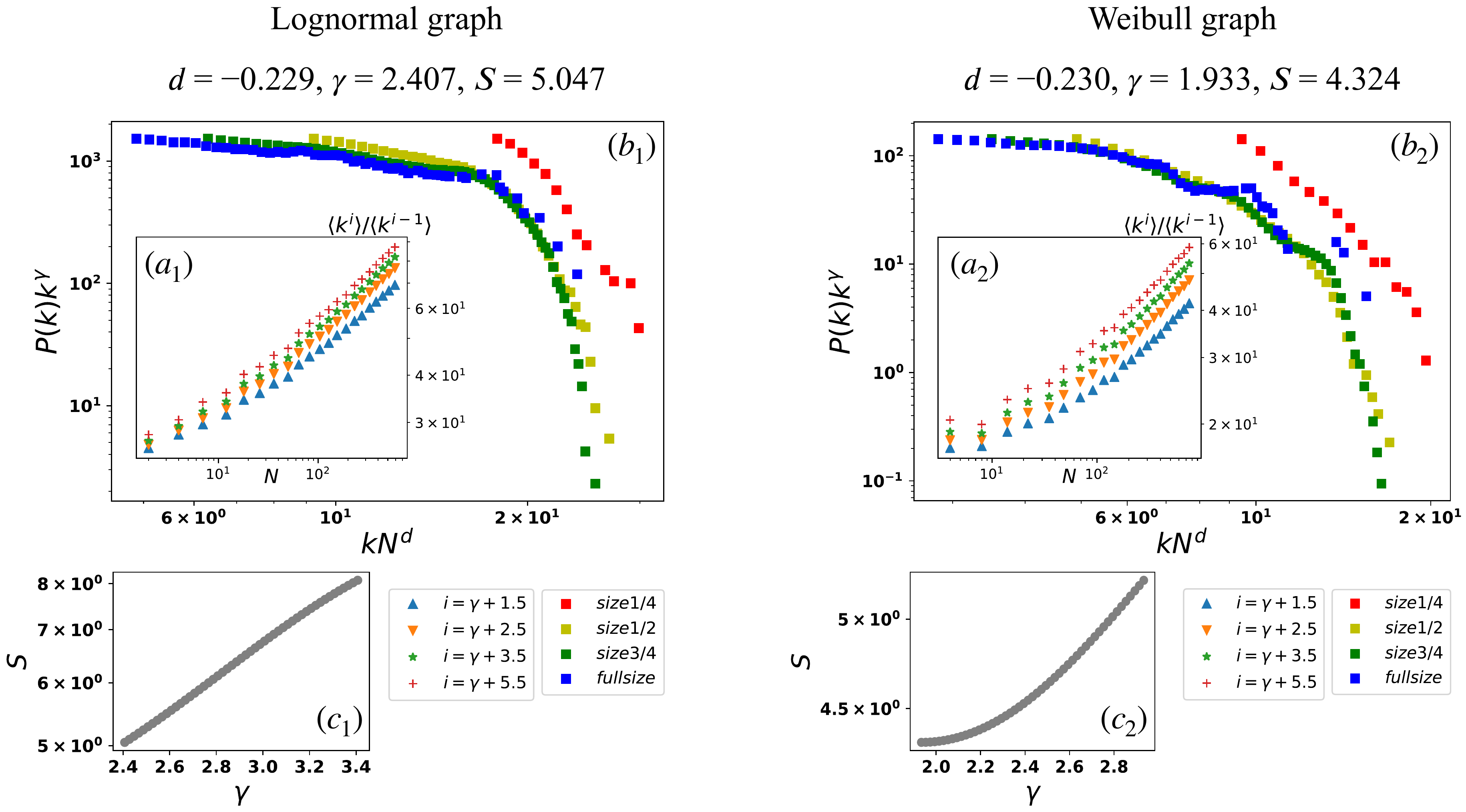}
\caption{Scaling analysis (with $N$) on a numerical realization of a log-normal graph with ($\sigma,\mu)=(0.8,1.8)$ (panels $a_1$, $b_1$, $c_1$) and of a Weibull graph with $(h,l)=(0.6,1.8)$ (panels $a_2$, $b_2$, $c_2$). In both cases the network has $N=10^4$ nodes. 
Log-normal graph: the best power law fit is obtained with $\Gamma=2.90\pm0.12$, the moment ratio tests yield $d=-0.209\pm0.033$ and $d_E=-0.208\pm0.033$, and the best collapse is obtained with $\gamma=2.40\pm0.37$ and yields $S=5.047$. 
Weibull graph: the best power law fit is obtained with $\Gamma=3.43\pm0.08$, the moment ratio tests yield $d=-0.230\pm0.037$ and $d_E=-0.219\pm0.036$, and the best collapse is obtained with $\gamma=1.933\pm1.055$ and yields $S=3.271$.}
\label{LogWei}
\end{center}
\end{figure*}

\begin{figure*}[p!] 
\begin{center}
\includegraphics[width=0.8\linewidth]{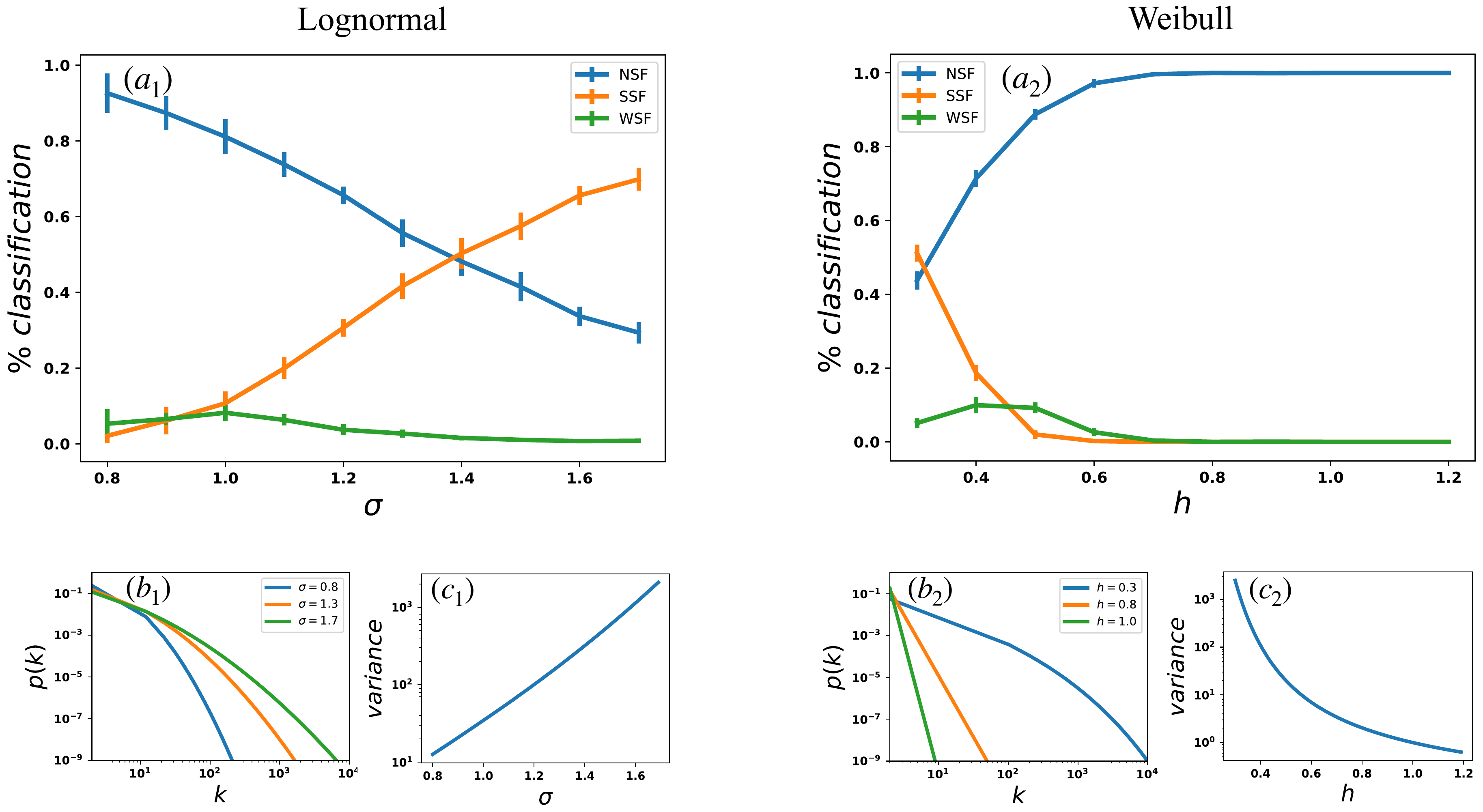}
\caption{Outcome of the scaling analysis (with $N$) on log-normal and Weibull networks as a function of the parameters of the degree distributions, respectively $(\mu,\sigma)$ and $(l,h)$. Panels ($a_1$) and ($a_2$) show the percentage of networks classified as strong, weak and non scale-free for varying $\sigma$ at fixed $\mu=1$, and for varying $h$ at fixed $l=3.5$, respectively. This statistics is computed over ensembles of 2000 networks for each choice of parameters $\sigma$ and $h$. Panels ($b_1$) and ($b_2$)  show representative instances of the distribution in the range of parameters analysed, whereas, panels ($c_1$) and ($c_2$) displays the corresponding value of the variance of the distribution. Note that we do not report results for varying $\mu$ at fixed $\sigma$ nor for varying $l$ at fixed $h$, because we observe almost no dependency of the classification on these parameters.}
\label{Logheat}
\end{center}
\end{figure*}

Figure \ref{LogWei} shows the scaling analysis for a realization of a network with log-normal $p(k)$ and for another realization with Weibull $p(k)$. In both cases we observe that the quality of the collapse is poor and that the moment ratios are not parallel lines. Therefore both networks are classified as NSF. Moreover, $S$ as a function of $\gamma$ does not show any minimum in the region around $\Gamma$ (the minimum does exist, but is located elsewhere). This means that the exponent estimated by finite size scaling $\gamma$ and that obtained from maximum likelihood power law fitting $\Gamma$ are substantially different: the outcome of the scaling analysis is not consistent in this case. 
However, the result depends much on the choices of parameters characterizing the distribution. Indeed Figure \ref{Logheat} shows that the percentage of networks classified NSF decreases by increasing $\sigma$ in the log-normal case, as well as by decreasing $h$ in the Weibull case -- up to a point where the variance of the distributions becomes so large that the scaling analysis can hardly distinguish these distributions from power laws at finite $N$. 
For these cases, the value of $\gamma$ that minimizes $S$ is indeed compatible with $\Gamma$.

\subsection*{Real world networks}

\begin{figure*}[ht!]
\begin{center}
\includegraphics[width=0.8\linewidth]{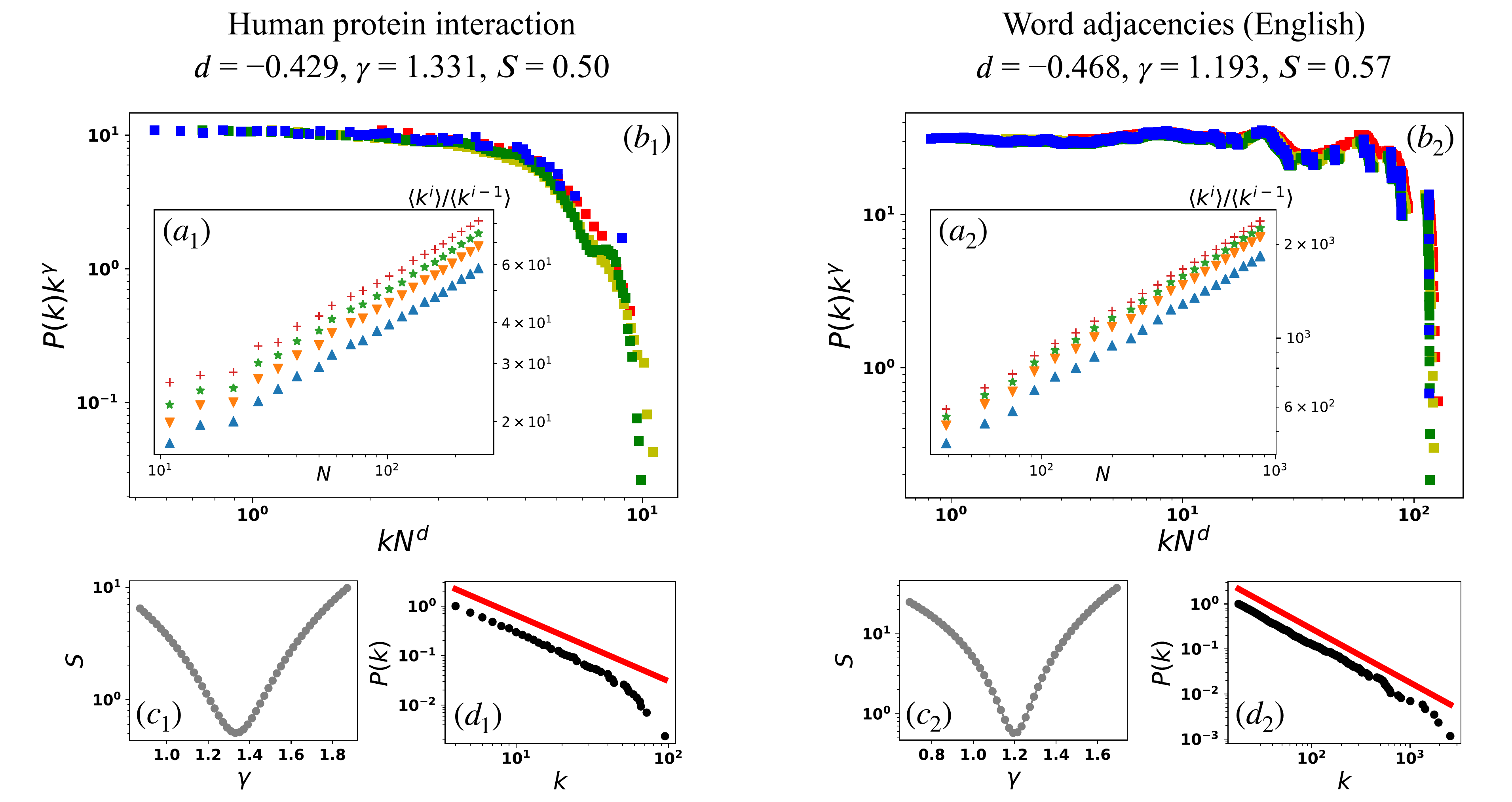}
\includegraphics[width=0.8\linewidth]{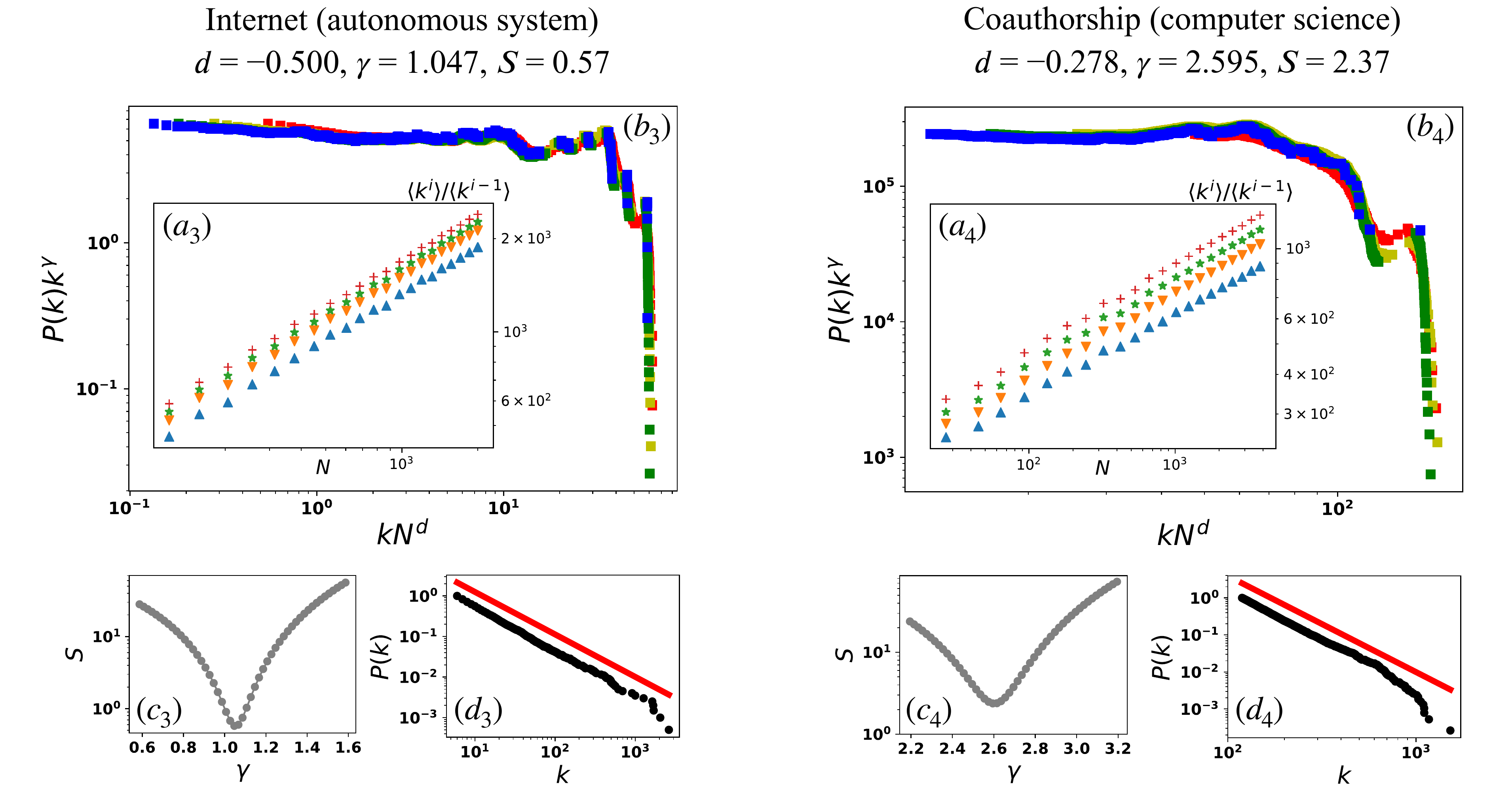}
\caption{Scaling analysis (with $N$) on four real network instances. 
Top left panels (1): the 2005 version of the proteome-scale map for Human binary protein-protein interactions ($N=1706$, $E=3155$) \cite{stelzl2005human}. 
Top right panels (2): the word adjacency graph extracted from the English text ``The Origin of Species'' by C. Darwin ($N=7724$, $E=46281$) \cite{milo2004superfamilies}. 
Bottom left panels (3): (symmetrized) snapshot of the Internet structure at the level of Autonomous Systems in 2007 ($N=26475$, $E=53381$) \cite{leskovec2007graph}. 
Bottom right panels (4): the collaboration graph of authors of scientific papers from DBLP computer science bibliography ($N=1314050$, $E=10724828$) \cite{10.1007/3-540-45735-6_1}. Panels (a), (b), (c) are analogous to those reported in Figures \ref{Barabasimodel} and \ref{LogWei}, whereas, panels (d) visually show the classical plots of $p(k)$ in double logaritmic scale together with the plot of the estimated slope $\gamma$ using FSS analysis.}
\label{realnets}
\end{center}
\end{figure*}

\begin{figure*}[ht!] 
\begin{center}
\includegraphics[width=0.8\linewidth]{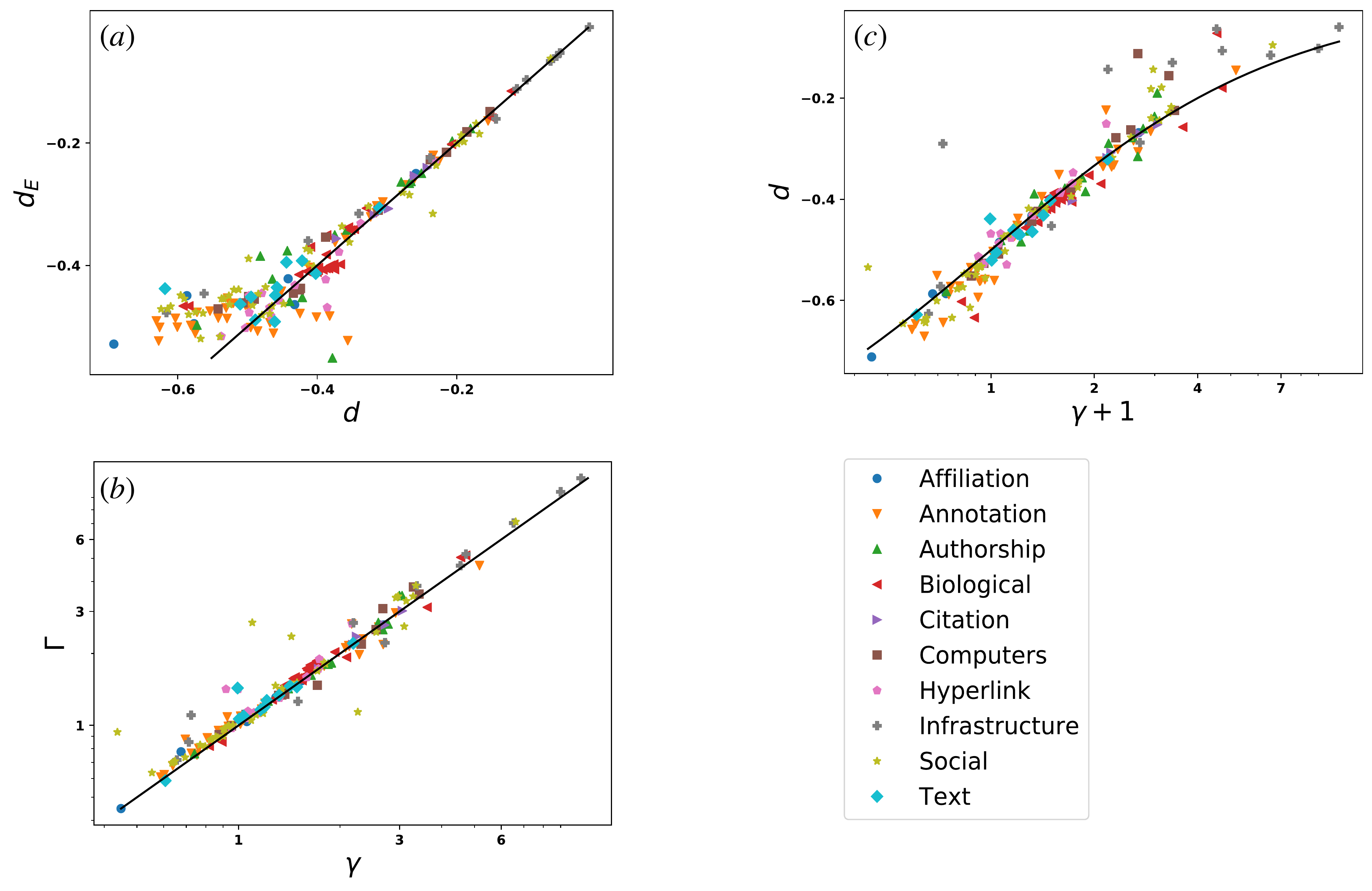}
  \caption{Visual summary of results from the finite size scaling analysis, in which  each network dataset is represented as a point in a specific plane. 
  Panel (a) shows the relation between $d$ and $d_E$ resulting from the moment ratio test, with the solid black line representing the identity. The other two panels refer to the scaling analysis with $N$. 
  Panel (b) shows the relation between $\gamma$ computed from finite size scaling and $\Gamma$ from the maximum likelihood power law fit of the degree distribution (see Methods). The solid line again represents the identity. 
  Panel (c) shows the relation between the exponents $\gamma$ and $d$ of the scaling function, with the solid black line representing the curve $d=-(\gamma+1)^{-1}$ (see the main text for details).}
  \label{Compare} 
\end{center}
\end{figure*}

At last we move to real network data. We consider a large set of empirical networks taken from the Index of Complex Networks (ICON) as well as from the Koblenz Network Collection (KONECT). These are the datasets used by Broido \& Clauset \cite{broido2018scale} and Voitalov {\em et al.} \cite{voitalov2018sfnwelldone}. See the Methods section for a discussion on how we built the dataset. Overall, we have networks belonging to ten different categories: biological (PPI), social (\ie, friendship and communication), affiliation, authorship (including co-authorship), citation, text (\ie, lexical), annotation (\ie, feature, folksonomy, rating), hyperlink, computer, infrastructure. Figure \ref{realnets} shows results of the finite size scaling analysis for selected network instances, whereas, Figure \ref{Compare} and Table \ref{tab:classification} summarize results of the scaling analysis for all the networks considered. The main outcomes of the analysis are the following.

\begin{itemize}
    \item Figure \ref{Compare}(a): the scaling exponents $d$ and $d_E$ obtained from the moment ratio test are compatible in most of the cases. 
    
    \item Figure \ref{Compare}(b): the value of $\gamma$ computed from finite size scaling is often in good agreement with $\Gamma$ obtained from the maximum likelihood power law fit of the degree distribution \cite{clauset2009power}.
    
    \item Figure \ref{Compare}(c): the exponents $\gamma$ and $d$ of the scaling function are not independent but satisfy a universal relation $d\simeq-(\gamma+1)^{-1}$, which derives from the nature of the degree cross-over in scale-free networks -- namely the maximum degree for which the power law behaviour holds. According to Eq. (\ref{scaling}), this is the value $k_{c}$ for which the scaling function $f(x)\to0$ 
    (graphically speaking, when the master curve $P(k)k^{\gamma}$ falls down), corresponding to $x\gtrsim 1$ whence $k_{c}\sim N^{-d}$. 
    The analysis presented in Figure \ref{Compare}(c) suggests that $k_{c}\sim N^{1/(\gamma+1)}$, and in agreement with theoretical results we find that also the maximum degree of the network $k_{max}$ scales in the same way (see Supplementary Information). However this scaling behavior is somehow different from the $k_{c}\sim N^{1/\gamma}$ as predicted by hand-waving argument \cite{dorogovtsev2002evolution,boguna2004cut,aiello2001random}, likely due to inner correlations in the networks which modify the value of the cross-over \cite{boguna2004cut}.

    \item No particular relation between quality of collapse $S$ and estimated exponent $\gamma$ is found, nor any clusterization of networks amenable to categories within the plane defined by these two variables (see Supplementary Information). However this result is obtained when the different network categories are well balanced in the dataset, because networks that are very similar tend instead to cluster together. This is for instance the case of protein interaction networks belonging to different species. In order to remove this artificial clustering effect, we have not considered in our dataset these (and other) cases of very similar networks nor repetitions of the same network (see Supplementary Information). This is the main reason why our dataset is apparently smaller than that used by Broido \& Clauset \cite{broido2018scale}.

    \item Overall, as shown in Table \ref{tab:classification}, the 185 networks of our dataset are classified as strong scale-free (SSF) in the 27\% of cases, weak scale-free (WSF) for the 23\% and non-scale-free (NSF) for 50\%. This classification however does vary substantially among the different network categories. On the one hand, biological networks are very often classified at least as WSF. The same happens for computer and hyperlink networks, with outliers respectively given by the Gnutella peer-to-peer file sharing network (that has the same character of a social networks \cite{wang2007analyzing}) and by some hyperlink networks restricted to specific domains. Citation and text networks are few in our analysis, but are often scale-free. On the other hand, infrastructure networks (\ie, road and flights network) are rarely scale-free (with the notable exception of Air traffic control systems), possibly because of the heavy cost of establishing a connection. Between these two extremes, there are the social and other kinds of networks (see for instance the well-known discussion of the Facebook case presented in \cite{5462078,ugander2011anatomy}, and that of other information sharing social network presented in \cite{zhou2011emergence}).
\end{itemize}

\begin{table*}[ht!]
  \centering
  \footnotesize
  \begin{tabular}{l|c|cccccccccc}
  	&	{\bf TOTAL}	&	Affiliation 	&	Annotation	&	Authorship	&	Biological	&	Citation 	&	Computer	&	Hyperlink 	&	Infrastructure 	&	Social	&	Text 	\\
  	\hline
number	&	185	    &	8    	&	38	    &	15  	&	30   	&	5	    &	13    	&	14   	&	12	    &	39	    &	11     	\\
{\bf SSF}	&	27\%	&	63\%	&	21\%	&	27\%	&	40\%	&	40\%	&	39\%	&	22\%	&	0\%	    &	13\%	    &	55\%	\\
{\bf WSF}	&	23\%	&	12\%	&	24\%	&	20\%	&	30\%	&	0\%	    &	38\%	&	21\%	&	17\%  	&	18\%	&	27\%	\\
{\bf NSF}	    &	50\%	&	25\%	&	55\%	&	53\%	&	30\%	&	60\%    &	23\%    &	57\%	&	83\%	&	69\%	&	18\%	   \\
\end{tabular}
  \caption{Classification of empirical networks (split into categories). For each category we report the total number of networks and the percentage of SSF, WSF and NSF instances. For detailed results on each network analyzed, see the Supplementary Dataset Table.}
  \label{tab:classification}
  \end{table*}

\section*{Discussion} \label{sec:IV}

Since the onset of network science, scale invariance of complex networks has been regarded as a universal feature present in real data \cite{faloutsos1999on,adamic1999power,caldarelli2000fractal,Goh12583,mislove2007measurement,clauset2009power} as well as reproduced in models  \cite{barabasi1999emergence,bianconi2001bose,caldarelli2002scalefree,dorogovtsev20004structure,medo2011temporal,mitzenmacher2004brief}. Thus the recent claim by Broido \& Clauset \cite{broido2018scale} that scale-free networks are rare created a stir, strengthening previous claims along the same direction \cite{PhysRevLett.94.168101,clauset2009power,Stumpf665}. Voitalov {\em et al.} \cite{voitalov2018sfnwelldone} replied to these arguments fitting data to generalized power laws, that is, regularly varying distributions $p(k)=l(k)k^{-\lambda}$ (where $l(k)$ is a function that varies slowly at infinity and thus does not affect the power law tail). By allowing deviations from the pure power law distribution at low $k$, they argued that scale-free networks are definitely not rare. Gerlach \& Altmann \cite{gerlach2019testing} very recently touched on this issue, showing that correlations present in the data can lead to false rejections of statistical laws when using standard maximum-likelihood recipes (in the case of networks, this can be important in the presence of degree-degree correlations).

In this work we go beyond statistical arguments and apply powerful tools from the study of critical phenomena in physics to analyse a wide range of model and empirical networks. Here we have showed that many of these networks spontaneously, without fine-tuning, satisfy the finite size scaling hypothesis, which, in turn, supports the claim that complex networks are inherently scale-free.

While a direct comparison with the results previously discussed would be interesting, the final results would be meaningless, given the differences in the underlying hypothesis of the different models. We showed how different hypothesis can lead to different results. The hypothesis underling our approach, which came from result previously obtained in the field of statistical mechanics and critical phenomena, goes beyond the applications they where initially thought for and it does not need the existence of a critical point. Together with the others, our methodology goes in the bag of tools a researcher can use in order to assess the scale freeness of a network. 

Our scaling analysis is based on the extraction of small representations of the networks using a random node selection scheme. Of course, an intrinsic limitation of any rescaling method applied to network data is the impossibility to consider system sizes spanning orders of magnitude. As a further general remark, finding a robust method to rescale (or coarse grain \cite{song2006origins,papadopoulos2012popularity}) a network is still an open issue in the literature since networks are not embedded in any Euclidean space. Commonly used approaches lack generality since they are based on the choice of the embedding geometric space \cite{garcia2018multiscale} or on the average path length \cite{song2005self}. In order to avoid {\em ad hoc} assumptions, we decided to follow the simplest (although not necessarily the most accurate) scheme. As shown in the Supplementary Information, by averaging over many extraction of the sub-network we are able to preserve the degree distribution of the original network, that is what we are interested in. 
Finally note our claims regards the self-similarity of the degree distribution, 
but we restrain ourselves in making general conclusions about the overall self-similarity of networks -- this would involve the study of other quantities such as clustering, average path length and so on  \cite{kim2007fractality}. 

\section*{Materials and Methods}

Here we report the steps to test the finite size scaling hypothesis of Eq. (\ref{scaling2}) together with the moments ratio test of Eq. (\ref{constant}). Note that in order to test Eqs. (\ref{scalingedge}) and (\ref{constant2}), one uses the number of edges $E$ ($e$) associated with each (sub-)network of size $N$ ($n$), and replaces $d$ with $d_E$.

\subsection*{Finite Size Scaling analysis}

Given an undirected network of size $N$, our analysis is based on the following steps.
\begin{enumerate}
\item We compute the degree distribution $p(k,N)$ and use the method of Clauset, Shalizi and Newman \cite{clauset2009power,alstott2014powerlaw} to estimate the best fitting power law parameters $\Gamma+1$ and $k_{min}$. 
\item We generate an ensemble of 100 sub-networks for each size $n\in\{\frac{N}{4},\frac{N}{2},\frac{3N}{4}\}$. Each sub-sample is obtained by picking $n$ nodes at random from the original network and by deleting all the other nodes and the links incident to them. We then compute the mean degree distribution $p(k,n)$ over each sub-network ensemble.
\item Both for the original network and for each sub-network, we check whether the (average) number of nodes $n^*$ with $k\ge k_{min}$ is larger than $\ln N$. If this condition is not met, we classify the network as non scale-free and the analysis ends. Otherwise, we proceed by removing the region below $k_{min}$ in both $p(k,N)$ and each $p(k,n)$, and renormalize them afterwards. As explained in the main text, this allows us to get rid of deviations at low degrees, including those induced by the sub-sampling (see also the Supplementary Information).
\item Using the moment ratio test, we determine $d$ (and its associated error) as follows. 
We compute a given moment ratio $\ang{k^i}/\ang{k^{i-1}}$ on each (sub-)network of size $n$, and use least-squares to fit  $\ln(\ang{k^i}/\ang{k^{i-1}})$ versus $\ln n$. We then average the resulting fit slope over different choices of the moments (indexed by $i$) to obtain $-d$. Note that since this test is computationally less expensive than the collapse analysis (see below), we use more than four sub-network sizes. In particular we use 20 equally spaced values of $n\in[\frac{N}{4},N]$, for each of which we compute the moments ratio (and associated error used as fit weight) over an ensemble of 100 $n$-sized sub-network built as described above. 
\item For each (sub-)network size  $n\in\{\frac{N}{4},\frac{N}{2},\frac{3N}{4},N\}$ we obtain the cumulative degree distribution $P(k,n)$. We then determine the exponents $\gamma$ and $d$ (and their associated errors) that maximizes the quality of the collapse plot (see below). Notably, the scaling exponent $d$ obtained from the collapse is always compatible with that obtained from the moment ratio test. Hence in order to decrease the computational cost of the method, one can in principle vary only $\gamma$ while keeping $d$ fixed at the value obtained from the moment ratios fit. 
\end{enumerate}

\subsection*{Quality of collapse} 

We now describe the procedure for deriving the master curve of the scaling function from the cumulative degree distributions of the various sub-networks, following the steps described in \cite{houdayer2004low,melchert2009autoscale}. The key premise is that when these distributions are properly rescaled they can be fitted by a single (master) curve. The quality of the collapse plot is then measured as the distance of the data from the master curve, and the collapse is good if all the rescaled distributions overlap onto each other. 

In practice for each (sub-)network size $n\in\{\frac{N}{4},\frac{N}{2},\frac{3N}{4},N\}$ we have the set $\{j\}$ of ordered points for the cumulative degree distribution in the form $\{(k_j,P(k_j,n))\}_j$. After applying the scaling laws we have:
\[
\begin{cases}
x_{nj}=k_j\,n^d\\
y_{nj}=P(k_j,n)\,k_j^{\gamma}
\end{cases}
\]
so that $x_{nj}$ is the rescaled $j^{th}$ degree in the distribution of the $n$-sized sub-network, and $y_{nj}$ is the rescaled value of such distribution relative to the $j^{th}$ degree. We also assign an error on the latter quantity as $dy_{nj}=dP(k_j,n)\,k_j^{\gamma}$, where $dP(k_j,n)$ is the Poisson error on the count $P(k_j,n)$ --- see the Supplementary Information. 

The master curve $Y$ is the function best fitting all these points. 
We define the quality of the collapse as
\begin{equation}
S=\cfrac{1}{3|M|}\sum_{(n,j)\in M}\cfrac{(y_{nj}-Y_{nj})^2}{dy_{nj}^2+dY_{nj}^2},
\end{equation} 
where $Y_{nj}$ and $dY_{nj}$ are the estimated position and standard error of the master curve at $x_{nj}$, while $M$ is the set of terms
of the sum (roughly, the set of points for which the curves for the various $n$ overlap).

For each $x_{nj}$, in order to define $Y_{nj}$ and $dY_{nj}$ we first need to select a set of points $m_{nj}$ as follows. In each of the other sets $n'\ne n$, we select (and put in $m_{nj}$) the two points $j'$ and $j'+1$ that best approximate $x_{nj}$ from below and above, \ie, the two points such that $x_{n'j'}\le x_{nj}\le x_{n'(j'+1)}$. If this procedure fails to select two points for each $n'\ne n$, then $Y_{nj}$ and $dY_{nj}$ are undefined at $x_{nj}$ which thus does not contribute to $S$ (this happens if set $n$ is alone in this region of $x$ and is the master curve by itself). Otherwise, we compute $Y_{nj}$ and $dY_{nj}$ using a linear fit through the selected points in $(n',l)\in m_{nj}$, so that $Y_{nj}$ is the value of that straight line at $x_{nj}$ and $dY_{nj}$ is the associated standard error:
\begin{equation}
Y_{nj}=\cfrac{ W_{xx} W_{y} - W_{x} W_{xy}}{\eta} + x_{nj}\cfrac{W W_{xy}-W_{x}W_{y}}{\eta}
\end{equation}
\begin{equation}
dY_{nj}^2=\cfrac{1}{\eta}(W_{xx}-2x_{nj}W_{x}+ x_{nj}^2W)
\end{equation}
where $w_{n'l}=1/dy_{n'l}^2$ for the fit weights and 
$W=\sum_{(n'l)\in m_{nj}} w_{n'l}$, 
$W_x=\sum_{(n'l)\in m_{nj}} w_{n'l}x_{n'l}$, 
$W_{y}=\sum_{(n'l)\in m_{nj}} w_{n'l}y_{n'l}$, 
$W_{xx}=\sum_{(n'l)\in m_{nj}} w_{n'l}x_{n'l}^2$, 
$W_{xy}=\sum_{(n'l)\in m_{nj}} w_{n'l}x_{n'l}y_{n'l}$, 
$\eta= W W_{xx}-W_{x}^2$ for the fit parameters. 

The  quality of the collapse $S$ measures the mean square distance of the sets to the master curve in units of standard errors, analogously to a $\chi^2$ test \cite{houdayer2004low}. The number of degrees of freedom can be estimated by noting that each of the $|M|$ points of the sum of $S$ has in turn 3 intrinsic degrees of freedom: $|m|$ points as described above (6 in our case) minus 2 from computing mean and variance of $Y$, minus 1. Hence by using $3|M|$ as normalization factor, $S$ should be around one if the data really collapse to a single curve and much larger otherwise. 

We optimize the quality $S$ of the collapse by varying the scaling exponents $\gamma$ in the interval $\Gamma-0.5 \leq \gamma \leq \Gamma+0.5$ and $d$ in the interval $d-0.1 \leq \gamma \leq d+0.1$.
The errors associated with $\gamma$ and $d$ are estimated with a $S+1$ analysis: $\Delta \gamma$ is such that $S(\gamma +\Delta \gamma)=S(\gamma)+1$ and $\Delta d$ is such that $S(d +\Delta d)=S(d)+1$.

\section*{Dataset}

We extract a collection of real network data from the Index of Complex Networks (ICON) at \url{https://icon.colorado.edu/} as well as the Koblenz Network Collection (KONECT) at \url{http://konect.uni-koblenz.de/}. The full list of networks we consider together with detailed results of the finite size scaling analysis are reported in the Supplementary Dataset Table. To define the dataset we select networks (removing duplicates appearing in both ICON and KONECT) according to the following criteria. 

First, to allow for a reliable scaling analysis, we only use networks with $N>1000$ and $E>1000$ (for computational reasons, we did not consider networks with more than 50 million links).
We then include undirected networks, as well as the undirected version of both directed and bipartite networks. Similarly, we consider binary networks as well as the binarized version of weighted and multi-edge networks. We however ignore networks that are marked as incomplete in the database. 
Importantly, among database entries that possibly represent the same real-world network we select only one (or at most a few) entry, and consistently we do the same for temporal networks (when there is only one snapshot, we ignore the time stamp of links). 

In practice, in KONECT we select only the Wikipedia-related networks in English language. For ICON the implications are more profound. We ignore interactomes of the same species extracted from different experiments, the (almost 100) fungal growth networks, the (more than 100) Norwegian boards of directors graphs, the (more than 100) CAIDA snapshots denoting autonomous system relationships on the Internet, networks of software function for Callgraphs and digital circuits ITC99 and ISCAS89. We consider only one instance of Gnutella peer-to-peer file sharing network, as well as a few instances of the (more than 50) within-college Facebook social networks and of the (about 50) US States road networks. Among the (more than 100) KEGG metabolic networks, we select 17 species trying to balance the different taxonomies. 

Thus, in our analysis, we do employ the same data source used by Broido \& Clauset \cite{broido2018scale}, but we avoid over-represented network instances. As explained in the main text, this procedure removes the clustering of similar networks shown in Figure \ref{realnets}, and leads to less biased conclusions on the scale-free nature of networks belonging to different categories.

\bigskip

\paragraph*{Acknowledgments.} GCi and GCa acknowledge support from the  European Project SoBigData++ (GA. 871042). 
AM acknowledges support from University of Padova through "Excellence Project 2018" of the Cariparo foundation.

\paragraph*{Author Contribution.} GCa conceived the experiment. MS and GCi performed the analyses of the dataset. AM coordinated the activities on finite size scaling analysis. All authors contributed to the interpretation of results, and to the writing of the manuscript.

%

\end{document}